\documentclass[11pt,a4wide]{article}
\usepackage{newlfont}
\usepackage{amsmath}
\usepackage{graphics}
\usepackage{float}
\usepackage{graphicx}
\usepackage{longtable}
\usepackage{epsfig}
\usepackage{multirow}
\usepackage{multicol}
\usepackage{verbatim}
\usepackage{pdflscape}
\usepackage{cite}
\usepackage{array}
\usepackage{longtable}     
\usepackage{ragged2e} 
\usepackage{tabularx}
\usepackage{soul}
\usepackage{color}

\begin{document}

\title{Charmonium spectrum and its decay properties}
\author{Zaki Ahmad\thanks{email: zakiahmad754@gmail.com; zaki.ahmad@gcwus.edu.pk} $^{2\dag}$, ~Ishrat Asghar\thanks{email: ishrat.asghar@ue.edu.pk}$^1$, Bilal Masud\thanks{email: bilalmasud.chep@pu.edu.pk}$^{2\dag}$, M. Atif Sultan\thanks{email: atifsultan.pu@gmail.com}$^{2\ddag}$$^{3\dag}$ \\
\textit{$\ast$Department of Physics, Govt. College Women University, Sialkot, Pakistan}\\
\textit{$^1$Department of Physics, University of Education Lahore, Faisalabad Campus, Faisalabad.}\\
\textit{$^{2\dag}$Centre for High Energy Physics, Univeristy of the Punjab, Lahore, Pakistan}\\
\textit{$^{3\dag}$School of Physics, Nankai University, Tianjin 300071, China}}
\date{}
\maketitle

\section*{Abstract}
In this work, we have calculated the mass spectrum, radiative decays, and strong decays of charmonium ($c\overline{c}$) by using the non-relativistic quark potential(NRQP) model. The wave functions are calculated by solving the radial Schr\"{o}dinger equation numerically, which are further used to compute the radiative decay widths of ($c\overline{c}$) states. The $^3P_0$ model is used to calculate the strong decay widths by using the simple harmonic (SHO) wave functions. The SHO parameter $\beta$ values for different $c\overline{c}$ states are calculated by fitting it to the numerical wave functions. We also compare our results with experimental data and other theoretically predicted results. We assign to the charmonium states X(3940), X(3872), X(3862), X(4350) likely quantum numbers of $\eta_c(3S)$, $\chi_1(2P)$, $\chi_0(2P)$ and $\chi_2(3P)$ states.
\section{Introduction}
Charmonium mesons are the bound states of charm quark and its anti-quark, {\em i.e.} $(c\bar{c})$, and their study is an effective tool to study the strong force in non-perturbative regime. During the past few years, great progress has been achieved in the observation of charmonium and charmonium-like states. The charmonium and charmonium-like states, such as $\chi_{c1}(3872)$, $\chi_{c0}(3915)$, $\chi_{c2}(3930)$, $\psi(4230)$, $\psi(4360)$, $\psi(4500)$, $\psi(4660)$ and $\psi(4700)$ were observed by CMS, CDF, LHCb, DO, BESIII, BABAR and Belle \cite{pdg-2022}. These observations have given a deeper understanding of the charmonium mesons and have resolved many puzzles. A better understanding of the charmonium spectrum can play a vital role to understand these recently observed states.

There have been many theoretical studies of the charmonium spectrum and decay widths using relativistic and non-relativistic quark potential models. In Ref.\cite{Man-2024}, the spectrum and open-charm strong decays of charmonium states are calculated in a coupled-channel model. The spectrum of charmonium mesons along with their radiative transitions and leptonic decays are studied in Ref.\cite{Kher-2018} using the Cornell potential model. In Ref.\cite{deng-2017}, the linear and screened potential model are used to predict the masses of charmonium mesons. The radiative transitions of some low lying states are also calculated in this work. The strong decay widths of charmonium mesons are calculated in Ref.\cite{Long-cheng-2018} using $^3P_0$ pair creation model. For $^3P_0$ calculations, the numerical wave functions are used for initial $c\bar{c}$ states and simple harmonic oscillator (SHO) wave functions are used for the final states. In Ref.\cite{zaki-2025} strong decay widths are calculated with two choices of wave functions: i) -numerical wave functions that are calculated using quark potential model with relativistic effects and ii) -simple harmonic oscillator (SHO) wave functions for both initial and final mesons.

In this paper, we calculate the masses, radiative and strong decay widths, along with the branching ratios, of $c\bar{c}$ mesons. Since charm quark is heavier, we use a NRQP model to predict the masses and wave functions of charmonium mesons. The spin-spin and spin-angular momentum interactions are also included in the Columbic plus linear potential. The parameters of the model are determined by fitting to the available experimental masses of the well-established $c\bar{c}$ states. These theoretical wave functions are further used to compute the radiative transitions of charmonium mesons. We use $^3P_0$ pair creation model to calculate the strong decay width of charmonium mesons. This $^3P_0$ model has been applied successfully for a variety of meson sectors~\cite{ackleh-1996,barnes-1997, barnes-2003, barnes-2005,close-2005, ferretti-2014, sun-2014, godfrey-2015, godfrey-2016-I, godfrey-2016-II,ferretti18, wang18, ishrat-2018, ishrat-2019, ishrat-2024}. The $^3P_0$ model depends on a parameter $\gamma$ \cite{ackleh-1996} which represents the pair creation probability. In this work, the parameter $\gamma$ is fitted by using the available strong decay data of charmonium states. We used simple harmonic oscillator (SHO) wave functions for initial and final mesons to calculate the strong decay widths. However, compared to Ref.\cite{Long-cheng-2018}, we fit $\beta$ parameter of SHO wave functions to the numerical wave functions obtained by solving radial Schr\"{o}dinger equation for initial and final mesons. In our earlier work \cite{ishrat-2018}, we noted that strong width is sensitive to the choice of SHO parameter $\beta$. Therefore, it is more realistic approach to use different $\beta$ values for each mesons. We combine radiative and strong widths to calculate the total widths and predict the branching ratios of all possible decay modes. These predictions are then used to identify X(3940), X(3872), X(3862), X(4350) states of charmonium mesons.

The organization of the paper is as follows. First, we describe the potential model used to calculate the mass spectrum of the charmonium, $D$ and $D_s$ mesons in Sec. (\ref{potential model}). E1 and M1 radiative transitions are calculated in Sec. (\ref{E1 and M1}). In Sec. (\ref{3p0 model}), we describe the $^3P_0$ decay model evaluation of strong decay widths using simple harmonic oscillator (SHO) wave functions. The results of these computations are compared with available experimental data and our results are given in Sec. (\ref{results}), while summary of this work is presented in Sec.(\ref{summary}).

\section{Potential model for quark antiquark}\label{potential model}
In this work, we use the following non-relativistic quark anti-quark effective potential \cite{barnes-2005}
\begin{eqnarray}
V_{q\overline{q}}(r) = -\frac{4\alpha_{s}}{3r} + b r + \frac{32\pi\alpha_{s}}{9m_{c}^2} (\frac{\sigma}{\sqrt{\pi}})^3 e^{(-\sigma^2 r^2)} \overrightarrow{S_q}.\overrightarrow{S_{\overline{q}}} + \frac{L(L+1)}{2 \mu r^2} + V_{spin-dep}+C,
\label{eq:pot}
\end{eqnarray}
where the first term describes the short distance one gluon exchange between quark and anti-quark, the second term describes the long distance confinement between quark and anti-quark, the third term represents the hyperfine interaction, and fourth is the centrifugal term. In equation (\ref{eq:pot}) $\alpha_s$ is the running coupling constant, $b$ is the string tension, $\sigma$ is a phenomenological parameter and $C$ is a phenomenological constant. The spin dependent term of equation  (\ref{eq:pot}) is
\begin{eqnarray}
V_{spin-dep}=\frac{1}{m_c^2}[(\frac{2\alpha_s}{r^3} - \frac{b}{r})\overrightarrow{L}.\overrightarrow{S} + \frac{4\alpha_s}{r^3}T],
\label{spin-dep}\end{eqnarray}
\begin{eqnarray}
\overrightarrow{L}.\overrightarrow{S}=\frac{J(J + 1) - L(L + 1) - S(S + 1)}{2},
\end{eqnarray}
\begin{equation}\label{triplet}
<^3L_J|T|^3L_J|>= \left\{
      \begin{array}{ll}
       -\frac{L}{6(2L + 3)}, $ $$ $$ $$ $$ $$ $J = L + 1 \\ \\
\frac{1}{6}, $ $$ $$ $$ $$ $$ $$ $$ $$ $$ $$ $$ $$ $$ $$ $J = L\\\\
-\frac{L(L + 1)}{6(2L - 1)}, $ $$ $$ $$ $$ $$ $J = L - 1.\\
      \end{array}
    \right.
\end{equation}
In this work, we have fitted the quark masses and the potential model parameters. We have taken the parameters for charmonium, $D$ and $D_s$ globally, and the fitted values of these parameters are reported in Table(\ref{gipar}). To find the mass spectrum of the $c\overline{c}$, we solved the radial Schr\"{o}dinger equation numerically. This equation is given as
\begin{eqnarray}
-\frac{1}{2\mu}U''(r) + V(r)U(r) = E U(r)\label{scheq1},
\end{eqnarray}
where $\mu$ is the reduced mass of the meson. Masses of mesons are found by using the following expression:
$$M=m_q+m_{\overline{q}}+E.$$
It is to be noted that the spin dependent potential is $\propto 1/r^3$. So, the Schr\"{o}dinger equation for $u(r\rightarrow 0)$ cannot be solved. In this work, we introduce a cutoff parameter within small range  $r\in (0,r_c)$, so that, as $r\rightarrow 0$, the potential $\propto 1/r_c^3$ which is finite constant \cite{deng-2017}. Our predicted masses of charmonium mesons are reported in Table (\ref{tab:masses}) and masses of $D$ and $D_s$ mesons are given in Table (\ref{tab:massesD}) and (\ref{tab:massesDs}).
\begin{table}[htpb]
\caption{Our Fitted Potential model parameters for $c\overline{c}$, $D$ and $D_s$ mesons.}
\label{gipar}
\begin{center}
\begin{tabular}{ccc}
\hline
\hline
Parameter             &  Value      \\
\hline
$m_c$                 &  1.63 GeV            \\
$m_{u/d}$             &  0.341 GeV            \\
$m_{s}$               &  0.473 GeV            \\
$\alpha$              &  0.70           \\
$b$                   &  0.10 GeV$^2$      \\
$\sigma$              &  0.95 GeV          \\
$C_{c\bar{c}}$        &  0.0 GeV        \\
$C_{D}$               &  -0.230 GeV        \\
$C_{D_s}$               &  -0.171 GeV        \\
\hline
\hline
\end{tabular}
\end{center}
\end{table}
\begin{longtable}{ccccccc}
\caption{\label{tab:masses} Experimental and theoretically calculated mass spectrum of $c\overline{c}$ states.} \\
\hline
  State  &  Our calculated mass   &  Expt.\cite{pdg-2022}& NR  \cite{barnes-2005} & GI \cite{godfrey1985mesons} & $\beta$   \\
        &       (GeV)                   & (MeV)                     &    (GeV)                      &   (GeV)    & (GeV)\\
\hline
\hline
$J/\psi(1 ^3S_1)$       & 3.1090  & $3096.9\pm0.006$  & 3.090    & 3.098  & 0.6645    \\
$\eta_c(1^1S_0)$        & 2.9877  & $2979.2\pm1.3$    & 2.982    & 2.975    & 0.7578    \\
$\psi^{'}(2^3S_1)$      & 3.6776  & $3686.10\pm0.06$  & 3.672    & 3.676    & 0.5056    \\
$\eta_{c}(2^1S_0)$      & 3.6462  & $3637.7\pm4.4$    & 3.630    & 3.623    & 0.5374    \\
$\psi(3^3S_1)$          & 4.0157  & $4040\pm10$       & 4.072    & 4.100    & 0.4246    \\
$\eta_{c}(3^1S_0)$      & 3.9956     &                   & 4.043    & 4.064 & 0.4388    \\
$\psi(4^3S_1)$          & 4.2875  & $4421\pm6$        & 4.406    & 4.450    & 0.3816    \\
$\eta_{c}(4^1S_0)$      & 4.2721  &                   & 4.384    & 4.425    & 0.3898    \\
$\psi(5^3S_1)$          & 4.5252  &                   &          &          & 0.3547    \\
$\eta_c(5^1S_0)$        & 4.5130  &                   &          &          & 0.3601     \\
$\psi(6^3S_1)$          & 4.7410  &                   &          &          & 0.3359     \\
$\eta_c(6^1S_0)$        & 4.7298  &                   &          &          & 0.3398    \\
$\chi_2(1^3P_2)$        &3.6024   & $3556.18\pm0.13$  & 3.556    & 3.550    & 0.4347    \\
$\chi_1(1^3P_1)$        &3.5621   & $3510.51\pm0.12$  & 3.505    & 3.510  & 0.4728    \\
$\chi_0(1^3P_0)$        &3.4151   & $3415.3\pm0.4$    & 3.424    & 3.445    &0.6306    \\
$h_c(1^1P_1)$           &3.5688   & $3525.41\pm0.16$  & 3.516    & 3.517    & 0.4666    \\
$\chi_2(2^3P_2)$        &3.9467   & $3927.2\pm0.26$   & 3.972    & 3.979    & 0.3963    \\
$\chi_1(2^3P_1)$        &3.9163   &                   & 3.925    & 3.953    & 0.4210    \\
$\chi_0(2^3P_0)$        &3.8185   & $3918.4 \pm 1.9$                  & 3.852    & 3.916    & 0.5072    \\
$h_c(2^1P_1)$           &3.9196   &  $3888.4 \pm 2.5$                 & 3.934    & 3.956    & 0.4172    \\
$\chi_2(3^3P_2)$        &4.2230   &                   & 4.317    & 4.337    & 0.3649    \\
$\chi_1(3^3P_1)$        &4.1965   &  $4286^{+8}_{-9}$                 & 4.271    & 4.317    & 0.3812    \\
$\chi_0(3^3P_0)$        &4.1161   &                   & 4.202    & 4.292    & 0.4322    \\
$h_c(3^1P_1)$           &4.1991  &                   & 4.279    & 4.318     & 0.3788    \\
$\chi_2(4^3P_2)$        &4.4641   &                   & 4.317    & 4.337    & 0.3433    \\
$\chi_1(4^3P_1)$        &4.4398   &                   & 4.271    & 4.317    & 0.3552    \\
$\chi_0(4^3P_0)$        &4.3689   &                   & 4.202    & 4.292    & 0.3896    \\
$h_c(4^1P_1)$           &4.4422  &                   & 4.279    & 4.318     & 0.3535    \\
$\chi_2(5^3P_2)$        &4.6826   &                   & 4.317    & 4.337    & 0.3275    \\
$\chi_1(5^3P_1)$        &4.6598   &                   & 4.271    & 4.317    & 0.3368    \\
$\chi_0(5^3P_0)$        &4.5952   &                   & 4.202    & 4.292    & 0.3621    \\
$h_c(5^1P_1)$           &4.6622  &                   & 4.279    & 4.318     & 0.3354    \\
$\psi_{3}(1^3D_3)$      & 3.8334  &  $3842.71\pm0.16\pm0.12$          & 3.806    & 3.849    & 0.3867    \\
$\psi_{2}(1^3D_2)$      & 3.8235  & $3823.7\pm 0.5$                  & 3.800    & 3.838    & 0.3983    \\
$\psi(1^3D_1)$          & 3.8038  & $3769.9\pm2.5$    & 3.785    & 3.819    & 0.4160    \\
$\eta_{c2}(1^1D_2)$     & 3.8243  &                   & 3.799    & 3.837    & 0.3964    \\
$\psi_{3}(2^3D_3)$      & 4.1212  &                   & 4.167    & 4.217    & 0.3651    \\
$\psi_{2}(2^3D_2)$      & 4.1116  &                   & 4.158    & 4.208    & 0.3739    \\
$\psi(2^3D_1)$          & 4.0932  & $4191\pm20$       & 4.142    & 4.194    & 0.3874    \\
$\eta_{c2}(2^1D_2)$     & 4.1119  &                   & 4.158    & 4.208    & 0.3727    \\
$\psi_{3}(3^3D_3)$      & 4.370   &                   &     &       & 0.3445            \\
$\psi_{2}(3^3D_2)$      & 4.3604  &                   &     &       & 0.3511            \\
$\psi(3^3D_1)$          & 4.3428  &                   &     &       & 0.3612            \\
$\eta_{c2}(3^1D_2)$     & 4.3605  &                   &     &       & 0.3504            \\
$\psi_{3}(4^3D_3)$      & 4.5941  &                   &     &       & 0.3288            \\
$\psi_{2}(4^3D_2)$      & 4.5847  &                   &     &       & 0.3341            \\
$\psi(4^3D_1)$          & 4.5675  &                   &     &       & 0.3420            \\
$\eta_{c2}(4^1D_2)$     & 4.5846  &                   &     &       & 0.3336            \\
$\chi_{4}(1^3F_4)$      & 4.0154  &                   &     &       & 0.3607           \\
$\chi_{3}(1^3F_3)$      & 4.0148  &                   &     &       & 0.3657            \\
$\chi_{2}(1^3F_2)$      & 4.0096  &                   &     &       & 0.3724            \\
$h_{c3}(1^1F_3)$        & 4.0140  &                   &     &       & 0.3650            \\
$\chi_{4}(2^3F_4)$      & 4.2723  &                   &     &       &0.3461             \\
$\chi_{3}(2^3F_3)$      & 4.2708  &                   &     &       & 0.3502            \\
$\chi_{2}(2^3F_2)$      & 4.2649  &                   &     &       & 0.3557             \\
$h_{c3}(2^1F_3)$        & 4.2701  &                   &     &       & 0.3496           \\
\hline
\footnotetext[1]{GI represents Godfrey-Isgur potential model \cite{godfrey1985mesons}}
\end{longtable}
\begin{table}[htb]
\center
\caption{\label{tab:massesD} Experimental and theoretically predicted mass spectrum of $D$ mesons.}
\begin{tabular}{ccccccc}
\hline
    State  &  Our calculated mass (GeV)   &  Expt. (MeV)\cite{pdg-2022}& $\beta$  \\
\hline
$D^*(1 ^3S_1)$       & 2.0299  & 2.00855   & 0.3406    \\
$D(1^1S_0)$        & 1.8596  & 1.86726   & 0.4295    \\

$D_2^*(1^3P_2)$        & 2.4407  & 2.46305   & 0.2732    \\
$D_{1}'(1^3P_1)$        & 2.4117  & 2.427   & 0.2942    \\
$D_0^*(1^3P_0)$        & 2.3238  & 2.3245   & 0.3361    \\
$D_{1}(1^1P_1)$           & 2.4167  & 2.422  & 0.2880    \\

\end{tabular}
\end{table}

\begin{table}[htb]
\center
\caption{\label{tab:massesDs} Experimental and theoretically predicted mass spectrum of $D_s$ mesons.}
\begin{tabular}{cccccc}
\hline
    State  &  Our calculated mass (GeV)   &  Expt. (MeV)\cite{pdg-2022}&  $\beta$  \\
\hline
$D_s^*(1 ^3S_1)$       & 2.1253  & 2.1122   & 0.3920    \\
$D_s(1^1S_0)$        & 1.9577  & 1.96834   & 0.4911    \\
$D_2^*(1^3P_2)$        & 2.5425  & 2.5691   & 0.3026    \\
$D_{1}'(1^3P_1)$        & 2.5031  & 2.53511   & 0.3233    \\
$D_0^*(1^3P_0)$        & 2.3180  & 2.3178   & 0.4558    \\
$D_{1}(1^1P_1)$           & 2.5128  & 2.4595   & 0.3220    \\
\end{tabular}
\end{table}
\section{Radiative Transitions}\label{E1 and M1}
The photon emission plays an important role to study a meson because it gives access to charmonium states with different quantum numbers $n, L, S$ and $J$ \cite{nosheen-2017}. The electric dipole (E1) and magnetic dipole (M1)are the leading order radiative transitions. For the E1 transitions we use the following expression \cite{barnes-2005}
\begin{equation}
\Gamma_{E1}(n^{2S+1}L_J\rightarrow n^{'2S'+1}L'_{j'}+\gamma)=\frac{4}{3}C_{fi}\delta_{SS'}e_c^2\alpha|<\psi_f|r|\psi_i>|^2E_\gamma^3\frac{E_f^{(c\overline{c})}}{M_i^{(c\overline{c})}},
\end{equation}
where $e_c$ is the electric charge of charm quark, $E_\gamma$ is the final photon energy, $E_f$ and $M_i$ are the final state energy and mass of initial state of charmonium respectively, $<\psi_f|r|\psi_i>$ is the matrix element including the final and initial radial wave functions and
\begin{equation}
C_{fi}=max(L,L')(2J'+1)\begin{Bmatrix} J & 1 &J' \\ L' & S & L \end{Bmatrix},
\end{equation}
where $J$ and $J'$ are the total angular momentum of initial and final mesons, $L$ and $L'$ are the orbital angular momentum of inital and final mesons, and $S$ is the spin of initial meson.
For the M1 transitions, we use the following expression \cite{barnes-2005}
\begin{equation}
\Gamma_{M1}(n^{2s+1}L_J\rightarrow n^{'2s'+1}L_J'+\gamma)=\frac{4}{3}\frac{2J'+1}{2L+1}\delta_{LL'}\delta_{S,S'\pm1}e_c^2\frac{\alpha}{m_c^2}|<\psi_f|\psi_i>|^2
E_\gamma^3\frac{E_f^{(c\overline{c})}}{M_i^{(c\overline{c})}}.
\end{equation}

\section{The $^3P_0$ model}\label{3p0 model}
\label{sect:open-flavor-strong-decays}
The strong widths of charmonium mesons above $D\overline{D}$ threshold are computed using the $^3P_0$ strong decay model. In this model, the open-flavor strong decay of a meson take place through the production of $q\overline{q}$ pair with a strength parameter that is determined by available experimental data~\cite{micu-1969}. This pair creation model has been applied successfully on a variety of meson sectors~\cite{ackleh-1996,barnes-1997, barnes-2003, barnes-2005,close-2005, ferretti-2014, sun-2014, godfrey-2015, godfrey-2016-I, godfrey-2016-II,ferretti18, wang18, ishrat-2018, ishrat-2019, ishrat-2024}.

\begin{figure}[h]
\center
\includegraphics[width=10cm]{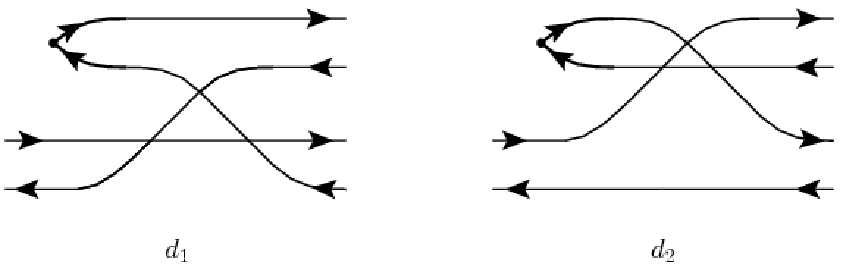}
\caption{Decay diagrams in $^3P_0$ Model.}
\label{3p0diagram}
\end{figure}

For strong decay process of meson  $A\rightarrow B+C$, the interaction Hamiltonian for the $^3P_0$ model is
\begin{equation}\label{hamiltonian}
H_I=2 m_q \gamma\int d^3\textbf{x}\;\overline{\psi}_q(\textbf{x}) \psi_q(\textbf{x}),
\end{equation}
where $\psi$ is the Dirac quark field and $\gamma$ is the pair-production strength parameter. Using the Fourier transformation of the above equation and taking the non-relativistic limit gives
\begin{equation}\label{eq: interaction-hamiltonian}
  H_I=2m_q\gamma \int d^3k[\overline{u}(\mathbf{k},s)v(\mathbf{-k},\overline{s})]b^{\dag}(\textbf{k},s)d^{\dag}(-\textbf{k},\overline{s}),
\end{equation}
where we have selected only the $b^{\dag}d^{\dag}$ pair creation term. Equation (\ref{eq: interaction-hamiltonian}) is used to calculate the matrix element $\langle BC|H_I|A\rangle$ for a process $A\rightarrow B+C$. There are two diagrams contributing in the matrix element, shown in Fig. (\ref{3p0diagram}). The flavor factor for each diagram along its multiplicity factor $\mathcal{F}$ are reported in Table (\ref{flavor factor}).

\begin{table}[h!]
\centering
\renewcommand{\arraystretch}{1.6}
\centering
\caption{Flavor factors for charmonium decay, where $|X_c\rangle\equiv|c\bar{c}\rangle$.}\label{flavor factor}
\begin{tabular}{c c c c c }
\hline
\hline
Generic Decay&Subprocess&$I_{flavor}(d_1)$&$I_{flavor}(d_2)$&$\mathcal{F}$\\
\hline\hline
$X_c\rightarrow D \bar{D} $        &  $X_c\rightarrow D^+ D^-$            & 0 & 1  & 2  \\
$X_c\rightarrow D^* \bar{D} $      &  $X_c\rightarrow D^{*+} D^-$         & 0 & 1  & 4  \\
$X_c\rightarrow D^* \bar{D}^* $    &  $X_c\rightarrow D^{*+} D^{*-}$      & 0 & 1  & 2  \\
$X_c\rightarrow D_s \bar{D}_s $    &  $X_c\rightarrow D_s^+ D_s^-$        & 0 & 1  & 1  \\
$X_c\rightarrow D_s^*\bar{D}_s $   &  $X_c\rightarrow D_s^{*+} D_s^-$     & 0 & 1  & 2  \\
$X_c\rightarrow D_s^*\bar{D}_s^* $ &  $X_c\rightarrow D_s^{*+} D_s^{*-}$  & 0 & 1  & 1  \\
\hline
\end{tabular}
\end{table}
The strong decay amplitude ($\mathcal{M}$) is calculated through the matrix element of both diagrams as:
\begin{equation}\label{decay-amplitude}
  \mathcal{M}_{LS}=\langle j_A,L_{BC},S_{BC}|BC\rangle \langle BC|H_I|A\rangle/\delta(\mathbf{A}-\mathbf{B}-\mathbf{C}).
\end{equation}
The above equation is combined with a relativistic phase space defined in Ref.~\cite{ackleh-1996} to compute the decay width of the process $A \rightarrow B+C$. The detail of the calculations of $\mathcal{M}$ by using the $^3P_0$ model is explained in our earlier work \cite{ishrat-2018,ishrat-2019}.\\
In this paper, we have computed strong decay widths of all the kinematically allowed open-flavor decay modes for each of the charmonium states mentioned in Table (\ref{tab:masses}) using the $^3P_0$ model. We use SHO wave functions for initial and final mesons which is given as
\begin{equation}\label{eq:sho}
  \psi_{nlm}(\textbf{k})=(2\pi)^{3/2}(-i)^{2n+l}N_{nlm}k^lY_{lm}(\hat{k})L_n^{l+1/2}(\beta^{-2}k^2)e^{-\frac{1}{2}\beta^{-2}k^2},
\end{equation}
where $\beta$ is the SHO parameter. In ref. \cite{barnes-2005}, a universal value of $\beta$ is taken for all initial and final mesons to calculate the strong widths of $c\bar{c}$. But in our work, we have taken different value of $\beta$ for each initial and final states: $\beta_A$ for initial meson, $\beta_B$ and $\beta_C$ for final mesons. In our work, the $\beta$ parameter is obtained by fitting the SHO wave function to the corresponding numerical wave function calculated by using the NRQP model. The sensitivity of strong decay width on SHO parameter $\beta$ has been explained briefly in our earlier work \cite{ishrat-2018}. Our calculated $\beta$ values for the initial charmonium mesons are reported in Table (\ref{tab:masses}) and for the final $D$ and $D_s$ mesons are mentioned in Table (\ref{tab:massesD}) and (\ref{tab:massesDs}) respectively.
\subsection{Paramater fitting of $^3P_0$ model}
We have determined the strength parameter $\gamma$  of the $^3P_0$ model by fitting it to
the experimental widths of five well known states: $\psi(3^3S_1)$, $\psi(4^3S_1)$, $\psi(1^3D_1)$, $\psi(2^3D_1)$ and $\chi_2(2^3P_2)$. Our fitted value of strength parameter $\gamma = 0.258$ well explains strong decay widths for these $c\bar{c}$ states. A comparison of our calculated strong decay widths with experimental decay widths of these experimental states is reported in Table (\ref{tab:3p0paramter}).
 \begin{table}[htb]
\center
\caption{\label{tab:3p0paramter} Strong decay widths of five well established $c\bar{c}$ states with fitted
value of strength parameter $\gamma=0.258$.}
\begin{tabular}{ccc}
\hline
    State  &  $\Gamma_{theo}$ (MeV)   &  $\Gamma_{exp}$ (MeV)\cite{pdg-2022}  \\
\hline

$\psi(3^3S_1)$   &  $70.98$ &$80\pm10$     \\
$\psi(4^3S_1)$   &  $72.10$   &$62\pm$20      \\
$\psi(1^3D_1)$   &   $27.93$   &$27.2\pm 1$     \\
$\psi(2^3D_1)$ &  $113.89$   &$70\pm10$\\
$\chi_2(2^3P_2)$   &  $26$      &$35.2\pm2.2$   \\
\hline
\end{tabular}
\end{table}
\section{Results and discussion} \label{results}
In this work, we use the non-relativistic potential model to solve the Schr\"{o}dinger equation numerically to find the wave function and spectrum of each $c\bar{c}$ state. We have calculated the charmonium masses up to 2F states. A comparison of our calculated masses and experimental masses are reported in Table (\ref{tab:masses}). Our predicted radiative transitions and strong decay widths of charmonoium states are reported in Tables (\ref{1s,2s and 3s decays}-\ref{1F decays}).\\
\par
Our calculated decay widths in Tables (\ref{1s,2s and 3s decays}-\ref{1F decays}) show that partial decay widths of M1 transitions are smaller than the corresponding E1 transitions due to the factors of $1/m_c$, and the branching ration ($Br$) of radiative transitions are large below the threshold $D\overline{D}$. These ratios decrease above the $D\overline{D}$ threshold due to the strong decay. Similar results have been observed in ref. \cite{Nosheen14,nosheen-2017}. \\
\par
1S and 2S states has only radiative transitions. Their strong decays are not allowed because their masses are below the $D\overline{D}$ threshold value. Our calculated masses of 1S and 2S states agree with experimental masses. These are reported in Table (\ref{tab:masses}). $\psi(2\;^3S_1)\rightarrow \eta_c(2^1 S_0) \gamma$ is the M1 transition.  Its predicted value agrees with the experimental value \cite{pdg-2022} and our calculated partial width M1 of $J/\psi(1\;^3S_1)\rightarrow \eta_c(1\;^1S_0)\gamma$ and $\psi(2\;^3S_1)\rightarrow \eta_c(1^1 S_0) \gamma$ is larger than the measured value \cite{pdg-2022}. E1 transition of $\psi(2\;^3S_1)\rightarrow \chi_{c0}(1^3 P_0) \gamma$ is close to the experimental value, and other E1 partial widths of $\psi(2\;^3S_1)$ are larger than measured values. Partial widths of E1 and M1 transitions of 1S and 2S states are given in Table (\ref{1s,2s and 3s decays}). \\
\par The first charmonium state above the threshold $D\overline{D}$ is $\eta_c(3\;^1S_0)$ which is allowed to decay strongly. Its mass is above threshold $D\overline{D}$. Our calculated mass is reported in Table (\ref{tab:masses}). The updated measured mass of the unknown charmonium like state $X(3940)$ is $3.942^{+7}_{-6}\pm6$ GeV and our calculated mass of $\eta_c(3S)$ is 3.9956 GeV, which is near to the $X(3940)$ $c\overline{c}$ state. So, we assign $X(3940)$ state to be the $\eta_c(3S)$ state. The experimental decay width of this state is $37^{+26}_{-15}\pm8$ MeV \cite{pdg-2022} and our predicted decay widths of this states is $112.82$ MeV, which is larger than the measured value. Ref.\cite{Long-cheng-2018} and \cite{barnes-2005} calculate decay widths of this state which are also larger than the experimental value. The partial decay width of $\eta_c(3S)$ is reported in Table (\ref{1s,2s and 3s decays}). The next 3S state is $\psi(3 ^3S_1)$. Its measured mass is $4.040\pm 10$ GeV and decay width of this state is $80\pm10$ MeV \cite{pdg-2022}. Our calculated mass of $\psi(3S)$ is $4.0157$ GeV and decay width is 70.98 MeV, which is agree with the PDG value. \\
\par
Our calculated mass 4S state of $\eta_c$  is $4.2721$ GeV, as given in Table (\ref{tab:masses}). Our calculated decay width of this state is 19.79 MeV, as reported in Table (\ref{4s decays}). $\eta_c(4S)$ has open charm strong decay channels as $DD^*$, $D^*D^*$, $DD_0^*$, $D_sD_s^*$, $D_s^*D_s^*$. Our predicted decay widths are quit different to other theoretical results \cite{Long-cheng-2018,barnes-2005}.\\
\par
The $\psi(4^3S_1)$ state has ten strong decay modes $DD$, $DD^*$, $D^*D^*$, $DD_1$, $DD'_1$, $DD_2^*$, $D^*D_0^*$, $D_sD_s$, $D_sD_s^*$, and $D_s^*D_s^*$. Its experimental measured width is $62\pm 20$ MeV \cite{pdg-2022} and our predicted decay width of this state is 72.10 MeV, which is agree with the experimental measured decay widths and other theoretical results \cite{Long-cheng-2018,barnes-2005}.\\
\par
The $\eta_c(5^1S_0)$ and $\psi(5^3S_1)$ have predicted mass as $4.5130$ GeV and $4.5252$ GeV, respectively. Our calculated decay widths of $\eta_c(5S)$ and $\psi(5S)$ states  are $102.88$ MeV and $92.35$ MeV, which are larger than other theoretical results \cite{Long-cheng-2018,barnes-2005}. \\
\par
The calculated decay widths of 1P states are reported in Table (\ref{1p and 2p decays}). The $\chi_{cJ}(1P)\rightarrow J/\psi(1S)$ and $\chi_{c2}(1P) \rightarrow h_c(1P)$ are below the threshold of $D\overline{D}$. So, their radiative widths are calculated. Our predicted partial widths of E1 and M1 transitions are in reasonable agreements with experimental measured values \cite{pdg-2022}, while E1 transition of $h_c(1P)\rightarrow \eta_c(1S)$ is in good agreement with the PDG average value \cite{pdg-2022}.\\
\par
The $\chi_2(2^3P_2$) has two open charm decay modes $DD$ and $DD^*$. Our calculated mass of this state is 3.9467 GeV, which is close to the experimental mass of 3.927 GeV \cite{pdg-2022}, and our measured decay width of this state is $35.2\pm2.2$ MeV. Our predicted decay with is 26 MeV, which is smaller than the experimental measured value \cite{pdg-2022}.\\
\par
The $\chi_1(2P)$ and $h_c(2P)$ have allowed open charm strong decay as $DD^*$. Their calculated decay widths are 88.71 MeV and 48.21 MeV, respectively. Our predicted decay width of $\chi_1(2^3P_1)$ is comparable with the Ref. \cite{Long-cheng-2018} result. The updated mass of the unknown charmonium state $X(3872)$ is $3871.69 \pm 0.17$ MeV and its measured decay width is less than 1.2 MeV \cite{pdg-2022}. Our calculated mass of this state is 3916.3 MeV, which is close to the $X(3872)$ state. So, it is a good candidate of $\chi_1(2P)$. The $\chi_0(2P)$ state decays into $DD$. Its predicted decay width is reported in Table (\ref{1p and 2p decays}). The measured mass of the unknown $c\overline{c}$ state X(3862) is $3888.4\pm 2.5 \text{ MeV}$, which is close to our calculated mass. So, the X(3862) is a good candidate of the $\chi_0(2P)$ state. Ref. \cite{L.Yu-2018,Wang-2022}  also assigns the same state to be $\chi_0(2P)$.\\
\par
The 3P charmonium states are not well established experimentally. We have calculated their radiative and strong decay widths. The charmonium like state X(4350) was observed and reported by Belle with $J^{PC}=2^{++}$ \cite{Belle-2010-x23p}. Its reported decay widths is $13^{+18}_{-9}\pm 4$ MeV \cite{pdg-2022} and we have assigned this state as $\chi_{c2}(3P)$. Our predicted decay widths is 22.17 MeV, which is comparable with the experimental value \cite{pdg-2022}. Ref. \cite{Long-cheng-2018,barnes-2005} also calculated the strong decay widths of this state as 43 MeV and 66 MeV, respectively, which is different to our predicted result. Our calculated masses and decay widths of 3P chamornium states are reported in Table (\ref{tab:masses}) and table (\ref{3p decays}). \\
\par
The $\psi_3(1^3D_3)$ and $\psi(1^3D_1)$ states above the threshold $D\overline{D}$, whereas $\psi_2(1^3D_2)$ and $\eta_{c2}(1^1D_2)$ are below the threshold. The decay mode of $\psi_3(1^3D_3)$ and $\psi(1^3D_1)$ is only $DD$. Our predicted total decay widths for  these are 1.29 MeV and 27.93 MeV, respectively. Our predicted partial E1 decay width of $\psi(1^3D_1)$ is very close the experimental decay width \cite{pdg-2022}. These are reported in Table (\ref{1D and 2D decays}). Our predicted total decay width of $\psi(1^3D_1)$ is in very good agreement with the measured decay width \cite{pdg-2022}. The $\psi_3(1^3D_2)$ and $\psi(1^1D_2)$ have E1 and M1 partial decay widths that are given in Table (\ref{1D and 2D decays}).\\
\par
The calculated masses and decay width of 2D states are reported in Table (\ref{tab:masses}) and (\ref{1D and 2D decays}), respectively. The $\psi(2^3D_1)$ state has five open charm decay modes $DD$, $DD^*$, $D^*D^*$, $D_sD_s$, and $D_s^*D_s^*$. Our calculated decay width is 113.89 MeV, which is larger than the measured decay width $70\pm10$ MeV \cite{pdg-2022} and agrees with the result $107\pm8$ MeV of the Crystal Ball and BESS Data \cite{kamalseth}.\\
\par
The 1F state is not well established experimentally. So, we compare our results with other theoretical findings \cite{barnes-2005}. Our predicted mass and decay width of 1F state are given in Table (\ref{tab:masses}) and (\ref{1F decays}). Our calculated decay width for the state $\chi_3(1^3F_3)$ is 67.14 MeV and $h_{c3}(1^1F_3)$ is 52.31 MeV has agreement with the results of ref.\cite{barnes-2005}, while $\chi_4(1^3F_4)$ and $\chi_2(1^3F_2)$ do not agree with the result of ref.\cite{barnes-2005}.\\

\section{Summary} \label{summary}
In this paper, we have studied the properties, including masses, radiative transitions, the Okubo-Zweig-Iizuka (OZI) allowed strong decays, total widths, and branching ratios, of charmonium mesons. Firstly, we calculated the mass spectrum of charmonium, $D$ and $D_s$ mesons using the NRQP model. The wave functions of quark anti-quark states are calculated by solving the radial Schr\"{o}dinger equation numerically. These numerical wave functions are used in computing the radiative and  widths of charmonium mesons. Strong decay widths of charmonium mesons above the $D\bar{D}$ threshold are calculated using $^3P_0$ pair creation model and fitted the SHO wavefunctions. The total decay widths of $c\bar{c}$ states are predicted by summing the radiative, and strong widths. The branching ratios of different final states are estimated by using the total widths. We assigned specific charmonium states on the basis of our predictions: $X(3940)$ is assigned to the $\eta_c(3S)$ state, $X(3872)$ to the $\chi_1(2P)$ state, $X(3862)$ to the $\chi_0(2P)$ state, and $X(4350)$ to the $\chi_2(3P)$ state. We also provide a comparison of our predicted masses and widths with available experimental data and recent theoretical studies.

\newpage
\Centering
\renewcommand{\arraystretch}{1.5} 
\setlength{\tabcolsep}{8pt}
\begin{longtable}{cp{3.0 cm}cp{3.0 cm} c p{2.0 cm}}
\caption{\label{1s,2s and 3s decays}E1, M1 radiative transition and strong decays of the 1S, 2S and 3S states.}\\
\hline\hline
Meson State&Decay Mode&$\Gamma_{exp}$ (keV)\cite{pdg-2022}& $\Gamma_{theo}$ (keV)&$Br(\%)$\\
\hline\hline
\endfirsthead
\multicolumn{4}{c}%
{\tablename\ \thetable\ -- \textit{Continued.}}\\
\hline\hline
Meson State&Decay Mode&$\Gamma_{exp}$(KeV)& $\Gamma_{theo}$ (KeV)&$Br(\%)$\\
\hline\hline
\endhead
\hline \multicolumn{4}{r}{\textit{Continued.}}\\
\endfoot
\endlastfoot
$J/\psi(1\;^3S_1)$  & $\eta_c(1\;^1S_0)\gamma$&1.7$\pm0.4$&2.33&100\\
\hline
$\eta_c(2\;^1S_0)$      &$h_c(1^1 P_1) \gamma$&&51.82&90.66\\
                        &$J/\psi(1^3 S_1)\gamma$&&5.34&9.34\\
                        &Total&&57.16&100\\
\hline
$\psi(2\;^3S_1)$        & $\chi_{c2}(1^3 P_2) \gamma$&26.64$\pm$1.7&54.31&39.78\\
                        & $\chi_{c1}(1^3 P_1) \gamma$&28.27$\pm$1.7&55.48&40.63\\
                        & $\chi_{c0}(1^3 P_0) \gamma$&29.57$\pm$1.6 &23.44&17.17\\
                        & $\eta_c(2^1 S_0) \gamma$&0.21$\pm$0.15 &0.175&0.13\\
                        & $\eta_c(1^1 S_0) \gamma$&1.24$\pm$0.29 &3.13&2.29\\
                        &Total&293$\pm$9&136.54&100\\
                        \hline
$\eta_c(3\;^1S_0)$     &$h_c(2^1 P_1)\gamma$&&43.34&0.0384\\
                        &$h_c(1^1 P_1) \gamma$&&16.97&0.0150\\
                        &$\psi(2^3 S_1) \gamma$&&0.70&0.00062\\
                        &$\psi(1^3 S_1) \gamma$&&4.21&0.0037\\
                        &$D D^*$     &&112.75\text{ MeV}& 99.94    \\
                        &Total       && 112.82 MeV &100\\
\hline
$\psi(3\;^3S_1)$        & $\chi_{c2}(2^3 P_2) \gamma$&&81.84&0.1153\\
                        & $\chi_{c1}(2^3 P_1) \gamma$&&51.94&0.0731\\
                        & $\chi_{c0}(2^3 P_0) \gamma$&&38.08&0.0536\\
                        & $\chi_{c2}(1^3 P_2) \gamma$&&16.75&0.0235\\
                        & $\chi_{c1}(1^3 P_1) \gamma$&&0.33&0.0005\\
                        & $\chi_{c0}(1^3 P_0) \gamma$&&7.18&0.0101\\
                        & $\eta_c(3^1 S_0) \gamma$&&0.14&0.0002\\
                        & $\eta_c(2^1 S_0) \gamma$&&0.38&0.0005\\
                        & $\eta_c(1^1 S_0) \gamma$&&2.38&0.0034\\
                        &DD                   &    &3.67 MeV& 5.17 \\
                          &DD$^*$                     && 8.73  MeV& 12.30   \\
                            &D$^*$D$^*$                 &         & 57.98MeV& 81.66  \\
                          &D$_s$D$_s$                 &            & 0.41MeV& 0.58 \\
                          &Total                   &$80\pm10$ MeV      &70.98 MeV&100        \\
                          \hline
\end{longtable}
\Centering
\renewcommand{\arraystretch}{1.5} 
\setlength{\tabcolsep}{8pt}
\begin{longtable}{cp{3.0 cm}cp{3.0 cm} c p{2.0 cm}}
\caption{\label{4s decays}E1, M1 radiative transition and strong decays of the 4S states.}\\
\hline\hline
Meson State&Decay Mode&$\Gamma_{exp}$ (keV)\cite{pdg-2022}& $\Gamma_{theo}$ (keV)&$Br(\%)$\\
\hline\hline
\endfirsthead
\multicolumn{4}{c}%
{\tablename\ \thetable\ -- \textit{Continued.}}\\
\hline\hline
Meson State&Decay Mode&$\Gamma_{exp}$(KeV)& $\Gamma_{theo}$ (KeV)&$Br(\%)$\\
\hline\hline
\endhead
\hline \multicolumn{4}{r}{\textit{Continued.}}\\
\endfoot
\endlastfoot
$\eta_c(4\;^1S_0)$     &$h_c(3^1 P_1) \gamma$&&64.37&0.3253\\
                        &$D D^*$       &&0.1 MeV& 0.5053\\
                        &$D^*D^*$     &&16.14 MeV&81.56\\
                        &$DD_0^*$       &&0.53 MeV&2.68\\
                        &$D_sD_s^*$      &&0.54 MeV&2.73\\
                        &$D_s^*D_s^*$   &&2.42 MeV&12.23\\
                        &Total     && 19.79 MeV&100\\
\hline
$\psi(4\;^3S_1)$          & $\chi_{c2}(3^3 P_2) \gamma$&&661.86&0.9180\\
                        & $\chi_{c1}(3^3 P_1) \gamma$&&468.18&0.6490\\
                        & $\chi_{c0}(3^3 P_0) \gamma$&&159.85&0.2217\\
                            &DD                          &         & 3.46 MeV&4.80\\
                           &DD$^*$                      &         & 14.86 MeV&20.61\\
                           &D$^*$D$^*$                  &         & 1.27 MeV&1.76\\
                           &DD$_1$                      &         & 23.82 MeV&33.04   \\
                           &DD$_1'$                     &         & 5.91 MeV& 8.20\\
                           &DD$_2^*$                    &         & 17.12 MeV&23.74\\
                           &D$^*$D$_0^*$                &         & 1.51 MeV&2.09\\
                           &D$_\text{s}$D$_\text{s}$    &         & 0.77 MeV&1.07\\
                           &D$_\text{s}$D$_\text{s}^*$  &         & 0.02 MeV&0.03\\
                           &D$_\text{s}^*$D$_\text{s}^*$ &         & 1.99 MeV&2.76\\
                           & Total  &$62\pm20$ MeV       &72.10 MeV &100\\
                           \hline
\hline
\end{longtable}
\Centering
\renewcommand{\arraystretch}{1.5} 
\setlength{\tabcolsep}{8pt}
\begin{longtable}{cp{3.0 cm}cp{3.0 cm} c p{2.0 cm}}
\caption{\label{5s decays}E1, M1 radiative transition and strong decays of the 5S states.}\\
\hline\hline
Meson State&Decay Mode&$\Gamma_{exp}$ (keV)\cite{pdg-2022}& $\Gamma_{theo}$ (keV)&$Br(\%)$\\
\hline\hline
\endfirsthead
\multicolumn{4}{c}%
{\tablename\ \thetable\ -- \textit{Continued.}}\\
\hline\hline
Meson State&Decay Mode&$\Gamma_{exp}$(KeV)& $\Gamma_{theo}$ (KeV)&$Br(\%)$\\
\hline\hline
\endhead
\hline \multicolumn{4}{r}{\textit{Continued.}}\\
\endfoot
\endlastfoot
$\eta_c(5\;^1S_0)$      &$h_c(4^1 P_1) \gamma$ &&85.78&0.0834\\
                        & $h_c(3^1 P_1) \gamma$&&11.20&0.0108\\
                        &$DD^*$       &&6.29 MeV&6.11\\
                        &$D^*D^*$       &&0.22 MeV&0.21\\
                        &$DD_0^*$        &&0.52 MeV&0.505\\
                        &$DD_2^*$       &&3.04 MeV&2.95\\
                        &$D^*D_1$        &&28.91 MeV&28.10\\
                        &$D^*D_1'$      &&32.59 MeV&31.68\\
                        &$D^*D_2^*$     &&31.02 MeV&30.15\\
                        &$D_sD_s^*$     &&0.05 MeV&0.05\\
                        &$D_s^*D_s^*$     &&0.11 MeV&0.11\\
                        &$D_sD_{s0}^*$ &&0.03 MeV&0.03\\
                             &Total    &&102.88 MeV&100\\
\hline
$\psi(5\;^3S_1)$    & $\chi_{c2}(4^3 P_2) \gamma$&&33.71 &0.0365\\
                    & $\chi_{c1}(4^3 P_1) \gamma$&&44.06 &0.0477\\
                    & $\chi_{c0}(4^3 P_0) \gamma$&&36.66 &0.0396\\
                    & $\chi_{c2}(3^3 P_2) \gamma$&&14.53 &0.0157\\
                    & $\chi_{c1}(3^3 P_1) \gamma$&&0.06 &0.00006\\
                    & $\chi_{c0}(3^3 P_0) \gamma$&&8.69 &0.0094\\
                    &$DD$             &&1.71 MeV&1.85\\
                    &$DD^*$          &&5.8 MeV&6.28\\
                    &$D^*D^*$        &&0.07 MeV&0.07\\
                    &$DD_1$          &&4.99 MeV&5.40\\
                    &$DD'_1$         &&2.35 MeV&2.54\\
                    &$DD_2^*$        &&2.16 MeV&2.34\\
                    &$D^*D_0^*$      &&0.61 MeV&0.66\\
                    &$D^*D_1$        && 14.19 MeV&15.36\\
                    &$D^*D'_1$       &&16.79 MeV&18.18 \\
                    &$D^*D_2^*$      &&42.41 MeV&45.92\\
                    &$D_sD_s$        &&0.43 MeV&0.46\\
                    &$D_sD_s^*$       &&0.13 MeV&0.14\\
                    &$D_s^*D_s^*$     &&0.17 MeV&0.18\\
                    &$D_sD_{s1}$      &&0.07 MeV&0.07\\
                    &$D_sD'_{s1}$      &&0.32 MeV&0.34\\
                    &$D_s^*D_{s0}^*$   &&0.01 MeV&0.01\\
                    &Total           &&92.35 MeV&100\\
\hline
\end{longtable}
\Centering
\renewcommand{\arraystretch}{1.5} 
\setlength{\tabcolsep}{8pt}
\begin{longtable}{cp{3.0 cm}cp{3.0 cm} c p{2.0 cm}}
\caption{\label{1p and 2p decays}E1, M1 radiative transition and strong decays of the 1P and 2P states.}\\
\hline\hline
Meson State&Decay Mode&$\Gamma_{exp}$ (keV)\cite{pdg-2022}& $\Gamma_{theo}$ (keV)&$Br(\%)$\\
\hline\hline
\endfirsthead
\multicolumn{4}{c}%
{\tablename\ \thetable\ -- \textit{Continued.}}\\
\hline\hline
Meson State&Decay Mode&$\Gamma_{exp}$(KeV)& $\Gamma_{theo}$ (KeV)&$Br(\%)$\\
\hline\hline
\endhead
\hline \multicolumn{4}{r}{\textit{Continued.}}\\
\endfoot
\endlastfoot

$\chi_{c2}(1^3 P_2)$    & $J/\psi(1^3 S_1) \gamma$  &370.6$\pm$34.6&306.65 &99.98 \\
                        & $h_c(1^1 P_1) \gamma$     &&0.05&0.016\\
                        & Total                          &&306.7&100\\
\hline
$\chi_{c1}(1^3 P_1)$    & $J/\psi(1^3 S_1) \gamma$  &284.8$\pm$23.6&249.94&100 \\
\hline
$\chi_{c0}(1^3 P_0)$    & $J/\psi(1^3 S_1) \gamma$  & 133.4$\pm$13.9&126.20&100\\
\hline
$h_{c}(1^1 P_1)$ & $\eta_c(1^1 S_0) \gamma$&357$\pm$246&368.53&99.84    \\
                 & $\chi_{c1}(1^3 P_1) \gamma$&&0.0054 & 0.0014\\
                 & $\chi_{c0}(1^3 P_0) \gamma$&&0.577& 0.1544\\
                 &Total                       &&369.11&100\\
                 \hline
$\chi_{c2}(2\;^3P_2)$   & $\psi(2^3 S_1) \gamma$ &&141.97& 0.5460  \\
                        & $J/\psi(1^3 S_1) \gamma$ && 139.92&0.5381 \\
                        & $\psi_3(1^3 D_3) \gamma$ &&19.38  & 0.0745\\
                        & $\psi_2(1^3 D_2) \gamma$ &&4.0 &0.0154\\
                        & $\psi(1^3 D_1) \gamma$  &&0.72 & 0.0028\\
                        & $h_c(1^1 P_1) \gamma$    && 0.55& 0.0021\\
                        &DD                         &              & 12.64 MeV&48.61\\
                          &DD$^*$                     &             & 12.49 MeV&48.04\\
                          &Total &$35.2\pm2.2$ MeV     &26                        &100 \\
\hline
$\chi_1(2\;^3P_1)$          & $\psi(2^3 S_1) \gamma$ &&162.05& 0.1826  \\
                        & $J/\psi(1^3 S_1) \gamma$&&76.01  &0.0857 \\
                        & $\psi_2(1^3 D_2) \gamma$ &&18.51& 0.0208\\
                        & $\psi(1^3 D_1) \gamma$&&19.12 &  0.0215  \\
                        & $h_c(1^1 P_1) \gamma$ &&0.02 &0.000022\\
                            &$D D^*$    &&88.43 MeV & 99.68\\
                            &Total        &&88.71 MeV&100\\

\hline
$\chi_0(2\;^3P_0)$      & $\psi(2^3 S_1) \gamma$ &&48.86& 0.0908  \\
                        & $J/\psi(1^3 S_1) \gamma$&&8.0&  0.0148  \\
                        & $\psi(1^3 D_1) \gamma$ &&4.83&   0.0089  \\
                        & $h_c(1^1 P_1) \gamma$ &&3.96& 0.0073\\
                        &$DD$      &&53.72 MeV & 99.86\\
                        &Total    &&53.79 MeV&100\\
\hline
$h_c(2\;^1P_1)$         & $\eta_c(2^1 S_0) \gamma$ &&216.02&0.4481  \\
                 & $\eta_c(1^1 S_0) \gamma$  &&165.20 & 0.3427\\
                 & $\eta_{c2}(1^1 D_2) \gamma$&&25.61 &0.0531\\
                 & $\chi_{c2}(1^3 P_2) \gamma$ &&0.82 & 0.0017\\
                 & $\chi_{c1}(1^3 P_1) \gamma$&& 0.02 &0.000041\\
                 & $\chi_{c0}(1^3 P_0) \gamma$ &&3.24 &0.0067\\
                        &$D D^*$   &&47.8 MeV  & 99.14\\
                        &Total&&48.21 MeV&100\\
\hline
\end{longtable}
\Centering
\renewcommand{\arraystretch}{1.5} 
\setlength{\tabcolsep}{8pt}
\begin{longtable}{cp{3.0 cm}cp{3.0 cm} c p{2.0 cm}}
\caption{\label{3p decays}E1, M1 radiative transition and strong decays of the 3P states.
.}\\
\hline\hline
Meson State&Decay Mode&$\Gamma_{exp}$ (keV)\cite{pdg-2022}& $\Gamma_{theo}$ (keV)&$Br(\%)$\\
\hline\hline
\endfirsthead
\multicolumn{4}{c}%
{\tablename\ \thetable\ -- \textit{Continued.}}\\
\hline\hline
Meson State&Decay Mode&$\Gamma_{exp}$(keV)& $\Gamma_{theo}$ (keV)&$Br(\%)$\\
\hline\hline
\endhead
\hline \multicolumn{4}{r}{\textit{Continued.}}\\
\endfoot
\endlastfoot
$\chi_2(3\;^3P_2)$      & $\psi(3^3 S_1) \gamma$ &&120.23&0.5423   \\
                        & $\psi(2^3 S_1) \gamma$ &&70.61  &0.0343      \\
                        & $J/\psi(1^3 S_1) \gamma$&&87.36  &  0.3940    \\
                        & $\psi_3(2^3 D_3) \gamma$ &&56.21 & 0.2535    \\
                        & $\psi_2(2^3 D_2) \gamma$ &&11.24 & 0.0506     \\
                        & $\psi(2^3 D_1) \gamma$ &&0.11 & 0.00049       \\
                        & $\psi_3(1^3 D_3) \gamma$ &&0.042 &  0.00018    \\
                        & $\psi_2(1^3 D_2) \gamma$ &&0.19&  0.00085     \\
                        & $\psi(1^3 D_1) \gamma$ &&0.062 &  0.00027      \\
                        & $h_c(1^1 P_1) \gamma$ &&0.68&0.0030
                        \\
                            & $D D$    &&2.27 MeV&10.23\\
                             &$D D^*$     &&13.11 MeV& 59.13 \\
                             &$D^*D^*$    &&5.94 MeV &26.79\\
                             &$D_sD_s$    &&0.46 MeV& 2.07\\
                             &$D_sD_s^*$  &&0.05 MeV&0.22\\
                             &Total &&22.17 MeV&  100\\
\hline
$\chi_1(3\;^3P_1)$          & $\psi(3^3 S_1) \gamma$&&103.34& 0.5681   \\
                        & $\psi(2^3 S_1) \gamma$ &&37.98   &  0.2087 \\
                        & $J/\psi(1^3 S_1) \gamma$ &&32.80 &  0.1803  \\
                        & $\psi_2(2^3 D_2) \gamma$ && 31.82& 0.1749 \\
                        & $\psi(2^3 D_1) \gamma$  &&0.77 &  0.0042 \\
                        & $\psi_2(1^3 D_2) \gamma$ &&0.67 &  0.0036  \\
                        & $\psi(1^3 D_1) \gamma$ &&0.069 &    0.00037  \\
                        & $h_c(1^1 P_1) \gamma$ && 0.014 &  0.000076\\
                            &$DD^*$    &&6.0 MeV&32.98\\
                             &$D^*D^*$   &&10.84 MeV&59.59\\
                             &$D_sD_s^*$ &&1.14 MeV&6.26\\

                             &Total      &&18.19 MeV&\\
\hline
$\chi_0(3\;^3P_0)$         & $\psi(3^3 S_1) \gamma$ &&19.86  &0.0272    \\
                        & $\psi(2^3 S_1) \gamma$ &&6.63    &  0.0091 \\
                        & $J/\psi(1^3 S_1) \gamma$ &&17.22 & 0.0236 \\
                        & $\psi(1^3 D_1) \gamma$ &&37.33 &  0.0512    \\
                        & $h_c(1^1 P_1) \gamma$ && 3.32 &  0.0045 \\
                            &$DD$       &&0.07 MeV&0.0961\\
                             &$D^*D^*$   &&71.9 MeV&98.75\\
                             &$D_sD_s$   &&0.76 MeV& 1.04 \\
                             &Total      &&72.81 MeV&100\\
\hline
$h_c(3\;^1P_1)$            & $\eta_c(3^1 S_0) \gamma$ &&172.58& 0.6881 \\
                 & $\eta_c(2^1 S_0) \gamma$ &&70.28 &  0.2802 \\
                 & $\eta_c(1^1 S_0) \gamma$ &&91.48 & 0.3647 \\
                 & $\eta_{c2}(2^1 D_2) \gamma$ &&45.36&0.1808 \\
                 & $\eta_{c2}(1^1 D_2) \gamma$ &&0.65& 0.0025\\
                 & $\chi_{c2}(1^3 P_2) \gamma$ &&0.82& 0.0032\\
                 & $\chi_{c1}(1^3 P_1) \gamma$ &&0.015&0.000059 \\
                 & $\chi_{c0}(1^3 P_0) \gamma$ &&3.71& 0.0148\\
                            &$DD^*$   &&10.43 MeV&41.58\\
                             &$D^*D^*$  &&13.3 MeV&53.03\\
                             &$D_sD_s^*$ &&0.97 MeV&3.87\\

                             &Total     &&25.08 MeV&100\\
\hline
\end{longtable}
\Centering
\renewcommand{\arraystretch}{1.5} 
\setlength{\tabcolsep}{8pt}
\begin{longtable}{cp{3.0 cm}cp{3.0 cm} c p{2.0 cm}}
\caption{\label{1D and 2D decays}E1, M1 radiative transition and strong decays of the 1D and 2D states.}\\
\hline\hline
Meson State&Decay Mode&$\Gamma_{exp}$ (keV)\cite{pdg-2022}& $\Gamma_{theo}$ (keV)&$Br(\%)$\\
\hline\hline
\endfirsthead
\multicolumn{4}{c}%
{\tablename\ \thetable\ -- \textit{Continued.}}\\
\hline\hline
Meson State&Decay Mode&$\Gamma_{exp}$(keV)& $\Gamma_{theo}$ (keV)&$Br(\%)$\\
\hline\hline
\endhead
\hline \multicolumn{4}{r}{\textit{Continued.}}\\
\endfoot
\endlastfoot
$\psi_3(1\;^3D_3)$   & $\chi_{c2}(1^3 P_2) \gamma$ &&403.73&31.24\\
                  & $\eta_{c2}(1^1 D_2) \gamma$&&0.001&7.7$\times 10^{-5}$ \\
                  &$D D$      &&0.88 MeV&68.21\\
                  &Total                        &&1.29 MeV&100\\
\hline
$\psi_2(1\;^3D_2)$        & $\chi_{c2}(1^3 P_2) \gamma$&&91.96&20.45\\
                  & $\chi_{c1}(1^3 P_1) \gamma$&&357.75 &79.55\\
                  &Total                        &&449.71&100\\
\hline
$\psi(1\;^3D_1)$        & $\chi_{c2}(1^3 P_2) \gamma$ &$<$17.41 & 5.35&0.0191  \\
                  & $\chi_{c1}(1^3 P_1) \gamma$ &81$\pm$ 27 & 118.15 &0.42\\
                  & $\chi_{c0}(1^3 P_0) \gamma$ &202$\pm$ 42 &189.86 &0.68\\
                        &DD                         &             &27.62 MeV&98.89 \\
                        &Total        &$27.2\pm1$ MeV          &27.93 MeV&100\\
                        \hline
$\eta_{c2}(1\;^1D_2)$       & $h_{c}(1^1 P_1) \gamma$ &&431.20 &100\\
                            \hline
$\psi_3(2\;^3D_3)$          & $\chi_{c2}(2^3 P_2) \gamma$&&264.50&0.51\\
                  & $\chi_{c2}(1^3 P_2) \gamma$&&36.29 &0.0713\\
                  & $\chi_{c4}(1^3 F_4) \gamma$&& 30.29&0.0595\\
                  & $\chi_{c3}(1^3 F_3) \gamma$&&2.44 &0.0048\\
                  & $\chi_{c2}(1^3 F_2) \gamma$&&110.08&0.2163 \\
                  & $\eta_{c2}(2^1 D_2) \gamma$ &&0.001&1.96$\times 10^{-6}$\\
                  & $\eta_{c2}(1^1 D_2) \gamma$&&0.049&0.000096 \\
                            &$D D$    &&0.06 MeV&0.001\\
                           &$D D^*$    &&18.77 MeV&36.88\\
                           &$D^* D^*$  &&31.36 MeV&61.62\\
                            &$D_s D_s$  &&0.18 MeV&0.35 \\
                            &$D_s D_s^*$&&0.08 MeV& 0.16\\
                           &Total      &&50.89 MeV&100\\
\hline
$\psi_2(2\;^3D_2)$         & $\chi_{c2}(2^3 P_2) \gamma$&&61.15&0.1087 \\
                  & $\chi_{c1}(2^3 P_1) \gamma$&&170.44 &0.3030\\
                  & $\chi_{c2}(1^3 P_2) \gamma$ &&3.21&0.0057\\
                  & $\chi_{c1}(1^3 P_1) \gamma$&&47.42& 0.0843\\
                  & $\chi_{c3}(1^3 F_3) \gamma$&&23.18& 0.0412\\
                  & $\chi_{c2}(1^3 F_2) \gamma$&&3.00 &0.0053\\
                  & $\eta_{c2}(1^1 D_2) \gamma$&&0.0017&  3.02 $\times 10^{-6}$\\
                            &$D D^*$   &&28.47 MeV&50.62\\
                           &$D^*D^*$     &&25.3 MeV&44.98\\
                           &$D_s D_s^*$  &&2.16 MeV&3.84\\
                           &Total        &&56.24 MeV&100  \\
\hline
$\psi(2\;^3D_1)$       & $\chi_{c2}(2^3 P_2) \gamma$ & &20.45&0.0179\\
                  & $\chi_{c1}(2^3 P_1) \gamma$ & & 280.22&0.2460\\
                  & $\chi_{c0}(2^3 P_0) \gamma$ &  &366.05& 0.3214\\
                  & $\chi_{c2}(1^3 P_2) \gamma$ &  &  0.00022& 1.93$\times 10^{-7}$  \\
                  & $\chi_{c1}(1^3 P_1) \gamma$ &  &  12.66 &0.0111\\
                  & $\chi_{c0}(1^3 P_0) \gamma$ &  & 154.36&0.1355\\
                    & $\chi_{c2}(1^3 F_2) \gamma$ &  & 173.09& 0.1519\\
                  & $\eta_{c2}(1^1 D_2) \gamma$ &  &0.36&0.0003\\
                        &DD                         &               & 31.71 MeV&27.84\\
                         &DD$^*$                     &              & 12.18 MeV&10.69\\
                         &D$^*$D$^*$                 &              & 62.89 MeV&55.22\\
                         &D$_\text{s}$D$_\text{s}$   &              &3.38 MeV&2.97\\
                         &D$_\text{s}$D$_\text{s}^*$ &              & 2.02 MeV&1.77\\
                         &Total              &$70\pm10$ MeV          &113.89 MeV&100 \\
                         \hline
$\eta_{c2}(2\;^1D_2)$       & $h_{c}(2^1 P_1) \gamma$&& 221.89&0.3541 \\
                     & $h_{c}(1^1 P_1) \gamma$ &&55.55 &0.0887\\
                     & $h_{c3}(1^1 F_3) \gamma$ &&26.76&0.0427\\
                     & $\psi_3(1^3 D_3) \gamma$ &&0.062&0.000098\\
                     & $\psi_2(1^3 D_2) \gamma$ &&0.0017&2.71$\times 10^{-7}$\\
                     & $\psi(1^3 D_1) \gamma$   &&0.156&0.0002\\
                            &$DD^*$    &&32.15 MeV&51.32 \\
                            &$D^*D^*$   &&28.78 MeV&45.94 \\
                             &$D_sD_s^*$ &&1.42 MeV&2.27 \\
                             &Total      &&62.65 MeV&100\\
                             \hline
\end{longtable}
\Centering
\renewcommand{\arraystretch}{1.5} 
\setlength{\tabcolsep}{8pt}
\begin{longtable}{cp{3.0 cm}cp{3.0 cm} c p{2.0 cm}}
\caption{\label{1F decays}E1, M1 radiative transition and strong decays of the 1F states.
.}\\
\hline\hline
Meson State&Decay Mode&$\Gamma_{exp}$ (keV)\cite{pdg-2022}& $\Gamma_{theo}$ (keV)&$Br(\%)$\\
\hline\hline
\endfirsthead
\multicolumn{4}{c}%
{\tablename\ \thetable\ -- \textit{Continued.}}\\
\hline\hline
Meson State&Decay Mode&$\Gamma_{exp}$(keV)& $\Gamma_{theo}$ (keV)&$Br(\%)$\\
\hline\hline
\endhead
\hline \multicolumn{4}{r}{\textit{Continued.}}\\
\endfoot
\endlastfoot
$\chi_4(1\;^3F_4)$     & $\psi_3(1^3D_3)\gamma$   &&232.72&1.53\\
                        &$h_c(1^1F_3)\gamma$   &&$4.46\times 10^{-6}$&2.94$\times 10^{-8}$\\
                        &$D D$     &   & 10.32 MeV &68.03\\
                            &$D D^*$    &  &  4.6 MeV& 30.32\\
                            &$D_s D_s$   & &   0.02 MeV&0.13\\
                            &Total       & & 15.17 MeV&100  \\
\hline
$\chi_3(1\;^3F_3)$  &$\psi_3(1^3D_3)\gamma$   &&25.70&0.038\\
                     &   $\psi_2(1^3D_2)\gamma$   &&224.84&0.33\\
                      &  $h_c(1^1F_3)\gamma$   &&$8.33\times 10^{-7}$&1.24$\times 10^{-9}$\\
                    &$D D^*$    &  & 67.16 MeV & 99.62\\
                    &Total      &  & 67.14 MeV& 100\\
\hline
$\chi_2(1\;^3F_2)$  &$\psi_3(1^3D_3)\gamma$   &&0.95&0.0010\\
                     &   $\psi_2(1^3D_2)\gamma$   &&36.53&0.0402\\
                     &   $\psi(1^3D_1)\gamma$   &&239.39&0.2637\\
                    &$D D$    &    & 45.88 MeV&50.55 \\
                   &$D D^*$    &  & 41.79  MeV&46.04\\
                   &$D_s D_s$  &  & 2.81  MeV&3.09\\
                   &Total       & &90.76 MeV &100 \\
\hline
$h_{c3}(1\;^1F_3)$  &$\eta_{c2}(1^1D_2)\gamma$&&252.78&0.48\\
                    &$D D^*$    &  & 52.05 MeV&99.50 \\
                   &Total      &  & 52.31 MeV&100  \\
\hline\hline
\end{longtable}
\label{results and discussion}

\end{document}